# The intriguing Flow Behavior of Soft Materials

Ranjini Bandyopadhyay*

*Department of Soft Condensed Matter, Raman Research Institute, CV Raman Avenue, Sadashivanagar, Bengaluru 560080, INDIA*

**Abstract**

Materials that can be deformed by thermal stresses at room temperature are called soft materials. Colloidal suspensions comprising solid particles evenly distributed in a fluid phase (smoke, fog, ink and milk, for example), emulsions(mayonnaise, lotions and creams), pastes (tomato ketchup, toothpaste), granular media (a bag of rice or sand), and polymer gels (polysaccharide gels) can be categorized as soft materials and are ubiquitous both at home and in industrial setups. Soft materials exhibit rich flow and deformation behaviors characterized by intriguing properties such as shear-thinning or thixotropy, shear-thickening or dilatancy, non-zero normal and yield stresses, *etc*. This article explains some of the mysterious flow properties of soft materials.

**What are soft materials?**

Several materials that we see around us every day such as lotions and creams, foodstuff such as jelly and mayonnaise, shaving foam, detergents, dyes and paints, *etc.,* are easily deformed by thermal fluctuations at room temperature. As diverse as these materials seem, they have some basic features in common, and are commonly referred to as soft materials or complex fluids. So, what are these common features that these seemingly different materials possess?

"What do we mean by soft matter? Americans prefer to call it 'complex fluids'. This is a rather ugly name which tends to discourage the young students. But it does indeed bring in two of the major features: *complexity* and *flexibility.*"These were the opening sentences of Professor Pierre Gilles de Gennes' Nobel lecture (Physics) on December 9,1991[1]. Soft materials are constituted by aggregates of molecules, or macromolecules, whose typical sizes are between 10 nanometers and 1 micrometer [2].Atomic length scales, in comparison, are of the order of angstroms. Owing to their larger sizes, macromolecules that make up soft materials move much more slowly than molecules in atomic systems. These properties endow soft materials with the complexity and flexibility that de Gennes told us about several years ago.

Another important consequence of the mesoscopic or intermediate sizes of the constituents of soft materials is that the inter-macromolecular interactions, which are dominated by thermal fluctuations, are of entropic origin. The shear modulus, or the ratio of shear stress to the shear strain[3], of a soft material is $G \sim \frac{k_B T}{a^3} \sim \frac{4.10^{-21} J}{(10^{-6})^3 m^3} \sim$ milli-Pascals (mPa) at room temperature. The moduli of conventional solids such as wood are of the order of gigapascals, while the modulus of stainless steel is approximately a few hundred gigapascals.

As a result of the weak inter-particle interactions (~ meV) that exist in these materials, small external perturbations can cause very large responses[4]. Examples of nonlinear behavior of soft materials therefore abound, for instance, the shear-dependent viscosity of a cornstarch suspension under an applied shear, pattern formation in an electrorheological fluid, *etc*. Indeed, these phenomena cannot be described by conventional Newtonian fluid mechanics[5]. Silly putty, for example, can be approximately modeled by assuming that its shear modulus is time-dependent, while shear thinning (decrease in viscosity under an applied shear) of colloidal suspensions such as blood or yogurt is described by modelling viscosity in terms of shear strain-rate dependent parameters.

Soft materials are characterized by rich out-of-equilibrium behavior[4]. Real-life examples include the formation of sand dunes (out-of-equilibrium pattern formation) and active dynamics (in living cells, such as in the transport of cargo along microtubules in a process driven by the hydrolysis of ATP). In this article, I shall introduce the reader to the intriguing flow and deformation behavior (rheology) of soft materials[5, 6].

I. **Rheology: the study of the flow and deformation of matter**

The flow of a soft material can be related to its structure and dynamics at the microscopic scale. Given the macromolecular nature of the constituents of soft materials, their structure and dynamics at the particle-scale are often studied using dynamic light scattering[7] or light microscopy[4]. Owing to the weak interactions that exist between constituent particles, small stresses can lead to dramatic macromolecular rearrangements and flow modifications in these materials[5, 6]. These flow modifications that happen at macroscopic length scales can be studied in the laboratory using a commercially available instrument called a rheometer[5].

The mesoscopic sizes of the constituent units and the weak inter particle forces in soft materials result in flow properties that are usually not exhibited by atomic solids. The time duration over which a soft material, once forced, relaxes to its equilibrium state typically ranges between a few 100 nanoseconds to as a long as a few seconds. This timescale is typically called the material's relaxation time.

Rheology is derived from the Greek words $\rho\epsilon o$ (reo-to flow) and $\lambda o\gamma\iota\alpha$ (logia-the study of). The term was coined by the American chemist Eugene Bingham and refers to the study of flow and deformation of matter. Historically, the mechanical behavior of materials under small deformations (linear rheology) was studied by 'rheologists', while that under complex deformations (nonlinear rheology) was studied by 'mechanicists'. In this article, I shall first introduce the reader to the simple ideas that are used to describe linear rheology. Finally, I shall give some examples of the intriguing flow properties of soft materials under large complex deformations (nonlinear rheology).

*a. Why do we study rheology?*

The study of flows is important in any industry where large-scale processes require the transport of materials. Many everyday objects such as lotions, foodstuff *etc.* can be characterized as soft materials. The understanding of flow and deformation is of great importance in material processing. That apart, soft materials are excellent candidates for scientists and engineers to study exciting and novel flow behaviors, particularly when the materials are sheared in their nonlinear rheological regimes. Specific observations will be discussed later in this article.

*b. Solids and liquids:*

Soft materials possess the unique property of *viscoelasticity*. Owing to the mesoscopic sizes and long relaxation times of the constituents when compared to atomic solids, soft materials exhibit solid-like (elastic) and liquid-like (viscous) properties at time scales accessible to us in the laboratory. Molecules of solids are tightly packed together which results in the material's *rigidity* and its *inability to flow*. Solids *store energy* and obey Hooke's Law which states that the stress response $\sigma$ of a solid material is directly proportional to the strain $\gamma$ applied on it: $\sigma = G\gamma$, where $G$ is the bulk/ rigidity/ elastic modulus[3]. A solid-like material is pictorially represented by a perfectly elastic spring[5, 6]. Liquids, in contrast, can flow and take the shape of the container they are confined in. Furthermore, liquids *dissipate* energy and their mechanical behavior can be represented by the Newton's law of viscosity: $\sigma = \eta\dot{\gamma}$ where $\eta$ is the shear viscosity of the liquid and $\dot{\gamma}$ is the rate at which the applied shear strain changes[8]. Liquids are represented by dashpots which are completely dissipative elements[5, 6].

*c. Viscoelastic materials*

As discussed in the last subsection, viscoelastic materials can simultaneously display solid-like and liquid-like properties. Soft materials are viscoelastic and are often pictorially represented by canonical combinations of springs and dashpots. The simplest canonical combinations are illustrated in Figure 1. Figures 1(a) and 1(c) present the Maxwell and Kelvin models of viscoelasticity, which are respectively represented by a series and a parallel configuration of one spring and one dashpot. In both the illustrations, the spring represents an element with elasticity $G$ while the dashpot represents an element with viscosity $\eta$. A fixed rate of strain $\bar{\dot{\gamma}}$, when applied to the Maxwell model [Figure 1(a)], gets distributed between the two elements depending on the relative values of the characteristic mechanical parameters, $\eta$ and $G$. As a direct consequence of the presence of the dissipative dashpot element, the stress-controlled Maxwell model is characterized by a stress decay over a relaxation time $\eta/G$ when the constant applied strain $\bar{\dot{\gamma}}$ is suddenly removed [Figure 1(b)]. This is in contrast to the instantaneous dissipation of energy in a Newtonian liquid upon the application of a shear rate $\dot{\gamma}$. Aqueous solutions of giant wormlike micelles[4] are known to display Maxwellian behavior. In the Kelvin-Voigt model [Figure 1(c)], the strain builds up exponentially with a characteristic relaxation time $\eta/G$ [Figure 1(d)] when a constant applied stress $\bar{\sigma}$ is applied to it. This behavior is in stark contrast to the instantaneously buildup of strain in a Hookean spring upon the application of a stress. The stress-controlled Kelvin-Voigt model can used to predict the *creep* behavior of viscoelastic solid-like materials. More complex rheological processes can be explained by using more complex arrangements of springs and dashpots[6].

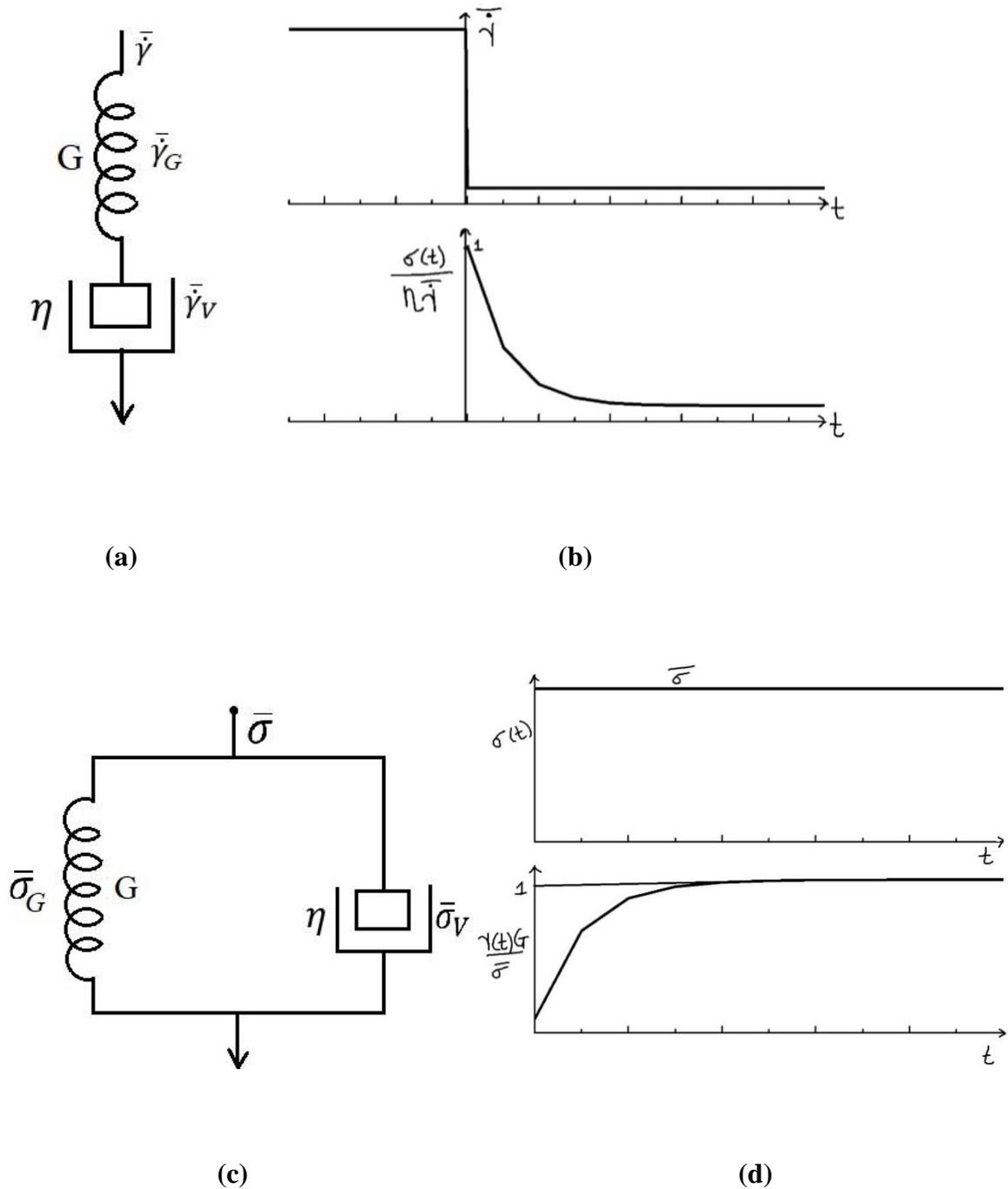

**Figure 1:** A schematic diagram showing the (a) Maxwell model with a dashpot (representing an element with viscosity $\eta$) and a spring (representing an element of elasticity $G$) connected in a series configuration. When a rate of shear strain $\bar{\dot{\gamma}}$ applied to this model is suddenly withdrawn, the stress $\sigma(t)$ decays exponentially as shown in (b). (c) presents an illustration of a Kelvin-Voigt model, where the spring and dashpot are connected in parallel. When a constant shear stress $\bar{\sigma}$ is applied, the strain $\gamma(t)$ in the Kelvin-Voigt model, as shown in (d), builds up exponentially. The Maxwell and Kelvin-Voigt models are the two simplest canonical models of viscoelasticity.

*d. Rheological responses*

Figure 2 shows examples of soft materials that we see around us every day. Many common food items (tomato sauce, jelly, mayonnaise) are viscoelastic[5, 6] and can be categorized as soft materials. The properties of these complex soft materials can be predicted to a limited extent by representing those using distinct combinations of springs and dashpots.

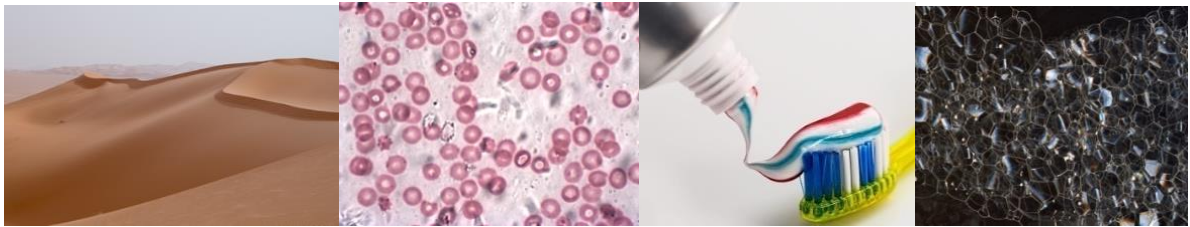

**Figure 2**: **Examples of everyday viscoelastic materials: from left to right -sand (granular matter) [9], blood (a colloidal suspension) [10], toothpaste (a gel) [11] and a soap foam (gas bubbles evenly distributed in a small amount of soapy liquid) [12]. The rheology of all these materials change dramatically due to the application of small stresses/ strains.**

Soft materials can have a range of responses. Figure 3 displays shear stress *vs.* shear strain rate plots that reveal a variety of mechanical responses that everyday viscoelastic materials display. In shear thinning or pseudo plastic behavior, often seen in blood, paint and ketchup, the viscosity (estimated from the slope of the shear stress *vs.* shear strain rate plot in Figure 3*) decreases* with increasing rate of shear. In dilatant or shear-thickening behavior, seen for example in suspensions of cornstarch particles subjected to appropriately high shear rates, the viscosity *increases* with increasing shear strain rate. In addition to these, two types of plastic responses are also illustrated in Figure 3. In contrast to shear thinning and shear thickening viscoelastic materials, plastic materials are characterized by the existence of a non-zero value of stress, the so-called yield stress, at a shear strain rate $\rightarrow 0$. Toothpastes and ketchups have non-zero yield stresses, which is why we have to squeeze the toothpaste tube or thump the ketchup bottle to make the contents flow. In a Bingham plastic, Hooke's law is obeyed once the yield stress is exceeded. In contrast, a Bingham pseudo plastic display shears thinning behavior under the same condition. A perfectly viscous or Newtonian response, where the shear stress is proportional to the rate of shear strain, is also illustrated in Figure 3 for comparison.

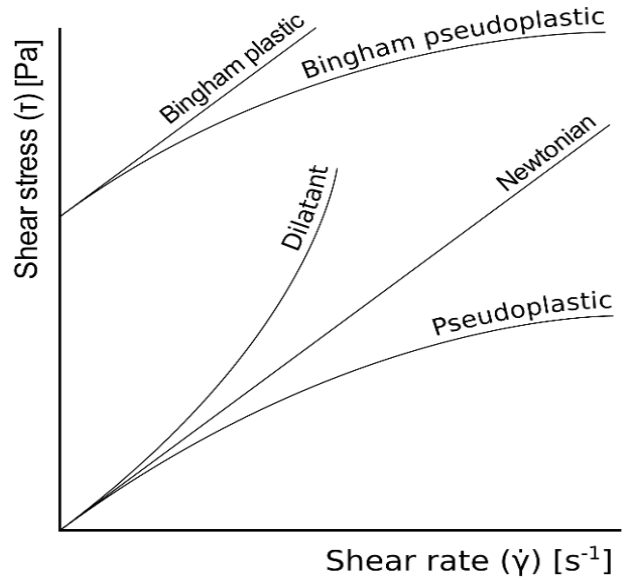

**Figure 3**: **Shear stress *vs*. shear strain rate plots for viscoelastic flows. Dilatants and pseudo plastic flows, commonly referred to as shear thickening and shear thinning respectively, and plastic flows are shown. Newtonian flow, where shear stress is directly proportional to shear strain, is also plotted.**

**Figure courtesy: https://upload.wikimedia.org/wikipedia/commons/e/e0/Non-Newtonian_fluid.svg**

*e. The superposition theorem and the theory of linear viscoelasticity*

The superposition principle for linear viscoelasticity proposes that the response of a material (its shear stress $\sigma(t)$) in the linear rheological regime is proportional to the initiating signal (the applied shear strain $\gamma(t)$). The general equation for linear viscoelasticity is written as a linear differential equation of the form:

$$\left(1 + \alpha_1 \frac{\partial}{\partial t} + \alpha_2 \frac{\partial^2}{\partial t^2} + \cdots + \alpha_n \frac{\partial^n}{\partial t^n}\right) \sigma(t) = \left(\beta_0 + \beta_1 \frac{\partial}{\partial t} + \beta_2 \frac{\partial^2}{\partial t^2} + \cdots + \beta_m \frac{\partial^m}{\partial t^m}\right) \gamma(t) \quad \text{--- (1)}$$

where *n, m* and the coefficients of the time differential terms ($\alpha_1, \ldots \ldots \ldots \beta_n$) are constants. In general, viscoelastic materials in the linear response regime are well described by Equation (1)[6]. It is easy to see that Equation (1) reduces to the Hooke's Law ($\sigma = E\gamma$) when $\beta_o = G$ (elastic modulus) $\neq 0$ and all other coefficients are 0. Similarly, Equation (1) reduces to the Newton's law of viscosity ($\sigma = \eta\dot{\gamma}$) when $\beta_1 = \eta$ (shear viscosity) $\neq 0$. Viscoelastic behavior ensues from Equation (1) when 2 or more coefficients are simultaneously non-zero. For example, the Kelvin-Voigt model of viscoelasticity (the pictorial depiction can be found in Figure 1(c)) is obtained if $\beta_o = G \neq 0$ and $\beta_1 = \eta \neq 0$ in Equation (1). Under these conditions, the equation reduces to

$$\sigma(t) = G\gamma + \eta\dot{\gamma}. \text{-----------------(2)}$$

If a constant stress $\bar{\sigma}$ is applied to the Kelvin-Voigt model at $t \geq 0$, the shear strain response, computed by solving Equation (2), shows a time-dependent buildup in shear strain $\gamma(t)$ that has been displayed in Figure 1(d) and can be described by the equation:

$$\gamma(t) = \frac{\bar{\sigma}}{G}[1 - \exp\left(-\frac{Gt}{\eta}\right)]. \text{----------(3)}$$

For modeling a range of viscoelastic materials using Equation (1), we consider different combinations of springs and dashpots by assigning appropriate non-zero values to the time-independent coefficients in Equation (1). Simpler canonical models involving only 2 elements, such as a spring and a dashpot in series or in parallel as shown in Figure 1, can be solved analytically. More complicated models are amenable only to numerical solution.

*f. Measurement of linear and non-linear rheology*

To understand the rheology of a material subjected to small deformations that lie within the linear rheological regime, the following experiments are performed to measure small strain relaxation functions[5, 6]:

i) **Stress relaxation experiments**: the stress relaxation function $G(t) = \frac{\sigma(t)}{\gamma_o}$ is computed by measuring the stress relaxation $\sigma(t)$ as a function of time $t$ at a constant step strain $\gamma_o$.

ii) **Creep and recovery measurements**: The time-dependent creep compliance $J(t) = \frac{\gamma(t)}{\sigma_o}$ is estimated by measuring the time dependent shear strain response $\gamma(t)$ under a constant applied step shear stress $\sigma_o$.

iii) **Oscillatory measurements:** The angular-frequency dependent elastic and viscous moduli, $G'(\omega)$ and $G''(\omega)$ respectively, are measured. This is achieved by measuring the amplitude of the oscillatory shear stress response $\sigma_o$ and the phase difference $\delta$ between the applied oscillatory shear strain $\gamma(t) = \gamma_o \sin(\omega t)$ and the oscillatory stress response $\sigma(t) = \sigma_o \sin(\omega t + \delta)$. A complex shear modulus $G^*$ is derived from the ratio of the measured oscillatory shear stress and the applied shear strain. The real part of $G^*$ is the elastic modulus $G'(\omega)$ while the imaginary part gives the viscous modulus $G''(\omega)$.

The small strain relaxation functions discussed above can be measured using an instrument called a rheometer. Images of an Anton Paar rheometer MCR 501 and some commonly used rheometer geometries are shown in Figure 4. Rheometers[4] can apply oscillatory and rotational shear strains or shear strain rates while simultaneously measuring shear stresses very accurately. The reverse experiment, wherein shear stresses are applied and shear strains/ shear strains rates are measured, can also be performed.

For higher strain values, it is difficult to compute relaxation functions using standard rheological models. The standard constitutive relations do not apply and a new treatment becomes necessary. Some theoretical models, such as those for second order fluids, have been proposed[5]. In spite of some advances, our understanding of flow and deformation properties of materials under large stresses or strains continues to be based largely on experimental observations. We will now briefly discuss some intriguing observations that arise in materials in the non-linear rheological regime.

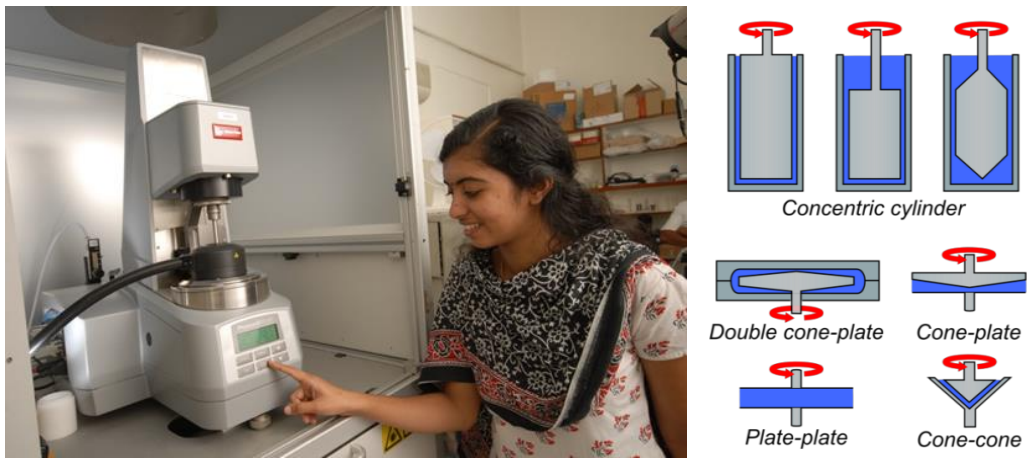

**Figure 4: (Left) An Anton Paar stress-controlled MCR 501 rheometer in the author's lab. In this picture, the rheometer is being operated by an ex-member of our laboratory, Dr. Paroor Harsha Mohan. (Right): Pictures of some commonly used rheometer geometries. Samples (shown in blue) are loaded in the geometry (usually made of stainless steel and depicted in grey). Oscillatory and rotational strains/ stresses are applied on the sample by the geometry and the stress/ strain responses are measured using a motor-transducer assembly.**

**Picture courtesy:**
**https://en.wikipedia.org/wiki/Rheometer#/media/File:Rotational_geometries.png**

*g. Nonlinear rheology of soft materials: some characteristic parameters and observations*

The nonlinear rheology of soft materials is characterized by non-zero normal stresses, shear thinning, and shear thickening etc.[5]. These properties, absent in conventional solids and liquids, arise due to rearrangements in the macromolecules constituting the microscopic structures of soft materials when large strains and stresses are applied. Below are some interesting observations that arise due to the complex nonlinear rheology of some common soft materials.

**i) Weissenberg Effect**: When a spinning rod is inserted in an elastic complex fluid (like a dense viscoelastic polymer solution), the sample, instead of being thrown outward as expected for a Newtonian fluid, moves towards and climbs the rod[5]. This is known as the Weissenberg effect, also referred to as the rod climbing effect. This effect arises due to positive normal stresses that characterize viscoelastic materials in flow. In contrast, a Newtonian liquid, when stirred, will move towards the edge of the container due to inertial forces. Figure 5 depicts the Weissenberg effect pictorially. Videos of rod climbing by viscoelastic fluids can be found at https://nnf.mit.edu/home/billboard/topic-5.

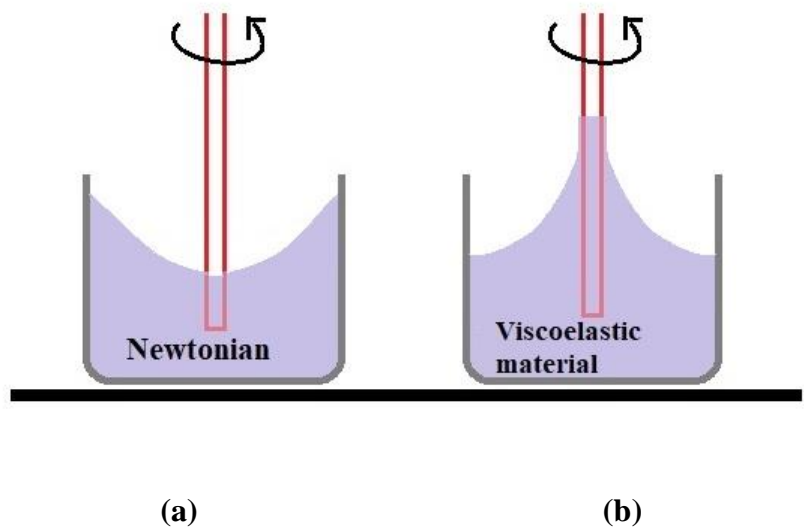

(a)          (b)

**Figure 5: (a) When a Newtonian liquid is stirred, the liquid interface is higher at the rims of the container than in the center. (b) A viscoelastic material, in contrast, climbs the spinning rod.**

**ii) Barus effect**: Dye swell, also known as the Barus effect, is seen while squeezing a dense soft material like an aqueous polymeric solution out of a small orifice[5]. The polymer swells immediately upon exiting the orifice. This can be understood by considering that polymers in a poor solvent like water prefer to exist as tiny globules. The polymers experience high shear rates during their passage through the orifice and are stretched out (shear thinning behavior) in the direction of their flow. Immediately after exiting the orifice, the polymers coil up into globules again and the material swells. This is depicted pictorially in Figure 6. A video of a viscoelastic polymer solution exhibiting the Barus effect can be found at https://www.youtube.com/watch?v=KcNWLIpv8gc.

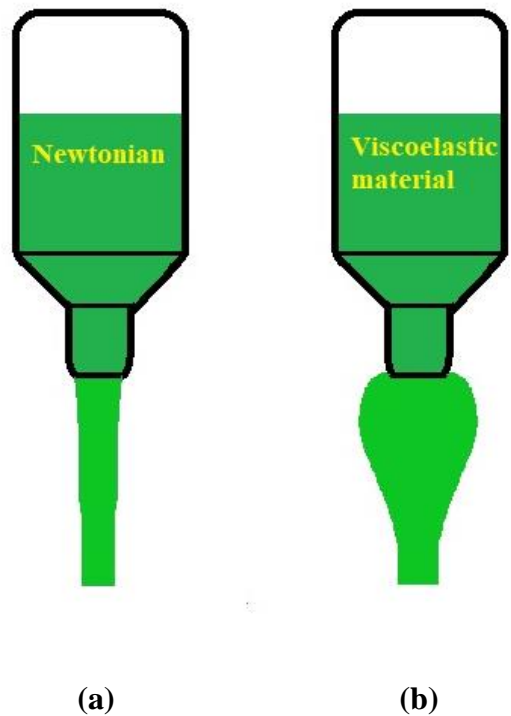

**(a)**          **(b)**

**Figure 6: (a) When a Newtonian liquid is squeezed out of an orifice, the diameter of the exiting liquid is the same as that of the orifice. In contrast, a viscoelastic material swells upon exiting the orifice as shown in (b).**

**iii) *Fano flow*:** Fano flow, also known as the tubeless syphon effect, is an observation that arises in elastic fluids due to their high extensional viscosity[5]. Extensional viscosity refers to the increase in a material's viscosity under extension. This property can be used to effectively syphon an elastic liquid out of its container using a syringe. Interestingly, the syringe can draw out the elastic liquid even without touching the surface of the latter once the siphoning initiated. A video of Fano Flow can be found at https://www.youtube.com/watch?v=aY7xiGQ-7iw.

**iv) *Kaye effect*:** An elastic liquid that is poured down an inclined plane shear thins as it slides down the plane. Thin jets of the elastic liquid are seen to spout out occasionally. This effect has been demonstrated even in shampoos and liquid soap. A video of the Kaye effect can be found at https://www.youtube.com/watch?v=wmUx-1o3Lzs.

v) *The bizarre flow of cornstarch suspensions*: A dense suspension of cornstarch particles behaves like a Newtonian liquid when stirred gently. On the other hand, the suspension solidifies if stirred vigorously. This counter-intuitive behavior arises due to the complex, non-monotonic flow behavior of cornstarch suspensions[13]. A nice video illustrating this phenomenon can be found at https://www.youtube.com/watch?v=Ja-6JtEZ7lk.

**Conclusions**

This article describes some interesting observations related to the flow of soft materials. On account of their sizes and relaxation times, soft materials are ideally suited for rheological measurements. The wide applicability of soft materials in our day to day lives and in industrial settings makes the study of their rheology extremely important. The interested reader may refer to a recent review by the author [14] for a more detailed monograph.

**Acknowledgment**

The author would like to acknowledge research funding from Raman Research Institute, Sadashivanagar, Bengaluru and Department of Science and Technology, Government of India, for funding *via* SERB grant EMR/2016/006757.